# Single-Phase High-Entropy Intermetallic Compounds (HEICs): Bridging High-Entropy Alloys and Ceramics

Naixie Zhou[1,2], Sicong Jiang[1], Timothy Huang[1], Mingde Qin[1], Tao Hu[3,1,*], Jian Luo[1,*]

[1]Department of Nanoengineering; Program of Materials Science and Engineering, University of California, San Diego, La Jolla, CA 92093-0448, USA

[2]Oerlikon Metco Inc., San Diego, CA 92121, USA

[3]State Key Laboratory of High Performance and Complex Manufacturing, School of Materials Science and Engineering, Central South University, Changsha, Hunan 410083, China

**Abstract:** High-entropy intermetallic compounds (HEICs) were fabricated by mechanical alloying and spark plasma sintering to fill a knowledge gap between the traditional high-entropy alloys (HEAs) and emerging high-entropy ceramics (HECs). Notably, several four- or five-component equimolar aluminides, such as the B2-phase $(Fe_{1/5}Co_{1/5}Ni_{1/5}Mn_{1/5}Cu_{1/5})Al$, have been made into single-phase HEICs for the first time. Thermodynamic modeling and a reversible, temperature-dependent, phase-stability experiment suggest that such B2-phase HEICs are entropy-stabilized phases. The structure of these HEICs resembles that of HECs with high-entropy mixing of four or five elements of nearly equal fractions in one and only one sublattice, but with significant (~10%) anti-site defects (differing from typical HECs). A new phase stability rule for forming single B2-phase HEICs is proposed. Five additional HEICs of predominantly $D0_{22}$ phases have also been made. This study broadens the families of equimolar, single-phase, high-entropy materials that have been successfully fabricated.

---

[*] Corresponding authors. E-mail addresses: jluo@alum.mit.edu (J.L.) and taohu1982@gmail.com (T.H.)

# 1. Introduction

High-entropy alloys (HEAs) consisted of at least four principal metallic elements with near equimolar fractions, also known as "multi-principal element alloys (MPEAs)" or "complex concentrated alloys (CCAs)," have attracted significant research interests in last 15 years [1-8]. Compared with traditional alloys with one primary element and several minor alloying dopants, HEAs explore new compositional spaces where no single component is dominant. Examples of HEAs include the famous FCC "Cantor alloy" CoCrFeMnNi [8] and refractory BCC HEAs (e.g., TaMoNbVW) [3, 4]. HEAs can often display unique mechanical properties, *e.g.* high strength and excellent cryogenic or high-temperature performance [1-7].

On the one hand, prior studies of metallic HEAs focus on simple solid solution phases where the multiple principal elements randomly occupy the same type lattice site of the FCC, BCC or HCP structure. On the other hand, an increasing number of new high-entropy ceramics (HECs) have been made in the last four years [9-19], where multiple principal metal cations occupy one sublattice, with another ordered anion sublattice with little or no mixing. Examples of single-phase HECs that have been successfully fabricated in last a few years include rocksalt oxides (e.g., $Mg_{0.2}Co_{0.2}Ni_{0.2}Cu_{0.2}Zn_{0.2}O$) [9, 20], metal diborides (e.g., $(Hf_{0.2}Zr_{0.2}Ta_{0.2}Nb_{0.2}Ti_{0.2})B_2$) [10, 21], fluorite oxides (e.g., $(Hf_{0.2}Zr_{0.2}Ce_{0.2}Y_{0.2}Gd_{0.2})O_{2-\delta}$) [12], perovskites (e.g., $(Ba_{0.5}Sr_{0.5})(Zr_{0.2}Sn_{0.2}Ti_{0.2}Hf_{0.2}Nb_{0.2})O_3$) [11, 13, 14], carbides (e.g., $(Hf_{0.2}Zr_{0.2}Ta_{0.2}Nb_{0.2}Ti_{0.2})C$) [15-18], and silicides (e.g., $(Mo_{0.2}Nb_{0.2}Ta_{0.2}Ti_{0.2}W_{0.2})Si_2$) [19]. These HECs also possess some unique or superior properties, such as reduced low thermal conductivities [16, 20] and increased hardness [15-18, 21].

This study further fills a knowledge gap between the traditional (metallic) HEAs and the emerging (mostly ionic) HECs via fabricating a new class of single-phase high-entropy intermetallic compounds (HEICs), exemplified by equimolar high-entropy aluminides such as the B2-phase $(Fe_{1/5}Co_{1/5}Ni_{1/5}Mn_{1/5}Cu_{1/5})Al$. These new HEICs are mostly metallic (albeit some mixed ionic-metallic bonds due to the different electronegativities) but have crystal structures resembling (mostly) ionic HECs: i.e., random mixing of four or five elements of equal molar fractions on one and only sublattice with another ordered sublattice with little mixing (albeit ~10% anti-site defects that are substantially higher than that in typical HECs). Thus, this discovery expands the families of single-phase, equimolar, high-entropy materials that have been successfully fabricated to date, and it bridges HEAs and HECs.

While the fabrication of single-phase HEICs has not been reported before, we recognize that multicomponent intermetallic compounds have been observed widely in complex alloys including

HEAs as secondary phases [6, 22-26]. Notably, Lu et al. reported an eutectic HEA, AlCoCrFeNi$_{2.1}$, and attributed the two phases as FCC and (ordered BCC-based) B2 based on X-ray diffraction (XRD) [22]. A follow-up study on the same AlCoCrFeNi$_{2.1}$ HEA by Nagase et al., however, suggested the formation of the ordered (FCC-based) L1$_2$ in the dendritic region, along with disordered FCC and BCC in the eutectic region, by using more sensitive electron diffraction [23]. Furthermore, four eutectic HEAs (CoCrFeNi$M_{0.45}$, where $M$ = Nb, Ta, Zr, or Hf) with FCC and Laves phases were designed and fabricated via casting [24]. These Laves phases are enriched in two or three elements (thereby being far away from equimolar compositions) [24]. CALPHAD modeling also suggested the existence of FCC-B2 two-phase regions in several HEAs [27]. Zhao et al. recently investigated three Al$_x$Co$_{0.2}$Cr$_{0.2}$Ni$_{0.2}$Ti$_{0.4-x}$ multi-phase HEAs with substantial amounts of multicomponent B2 phases that presumably have three principal elements (Co, Cr, and Ni) on one sublattice and another two (Al and Ti) primarily on the other sublattice [26]; this represents perhaps one reported case that is the most close to (but not yet truly) HEICs (and not single-phase). Moreover, none of these prior studies aimed at fabricating and investigating single-phase HEICs. Interestingly, two recent studies by Yang et al. [6] and He et al. [25] used multicomponent intermetallic nanoparticles (MCINPs) of the L1$_2$ [6] and D0$_{22}$ [25] structure, in the form of nanoscale precipitates in the continuous matrix of complex alloys [6] or HEAs [25], to strengthen the FCC-based alloys. Here, these MCINPs, e.g., (Ni$_{35}$Co$_{17}$Fe$_8$)(Al$_7$Ti$_{11}$Co$_2$) L1$_2$ (resemble Ni$_3$Ti or Ni$_3$Al) [6] and Ni$_{65.2}$Nb$_{24.1}$Co$_{7.7}$Cr$_{1.4}$Fe$_{1.3}$ D0$_{22}$ (close to Ni$_3$Nb) phases, are enriched in only one, or at most two, principal element(s) in each sublattice (so they are not multi-principal element high-entropy phases), and they are not the primary phases. Thus, the current study aims to make the first research effort to fabricate and subsequently characterize and investigate single-phase HEICs.

To seek the existence of single-phase HEICs with random mixing of four or five elements of a nominally equimolar fraction on one and only one sublattice (resembling HECs), we selected B2 phase high-entropy aluminides as our primary model systems. In addition, we further explored D0$_{22}$-phase high-entropy aluminides as a second structure to extend generality of this study. Aluminides were chosen in this study because they are important structural materials for their light weight, excellent thermal stability, outstanding oxidation resistance, and high strength [28]. The B2 phase is an ordered BCC-based structure, i.e. the CsCl type structure albeit significant anti-site defects. Fig. 1(a) shows the schematic structure of a B2-phase high-entropy aluminide, in which four or five elements of equimolar fractions are randomly mixed on one and only one sublattice and Al atoms primarily occupy the other sublattice. In this study, we fabricated seven equimolar B2-phase high-entropy aluminides, including three with single (and others with predominant one) ordered B2 phases. We also showed that these B2-phase HEICs are likely entropy-stabilized phases, and we further proposed a new selection criterion for forming single-phase B2 high-entropy aluminides (as a new phase stability rule). Furthermore, we successfully fabricated five additional high-entropy aluminides of primarily $D0_{22}$ phases (with the structure illustrated in Fig. 1(b)) to extend the generality of this study and our discovery.

## 2. Materials and Methods

To prepare the B2-phase high-entropy aluminides, we designed seven equal-aluminide-molar compositions (as listed in Table 1). Each non-Al element has equal molar fraction and the total molar fraction of these elements is 50% to maintain the stoichiometry of aluminide. The elements were selected based on the phase stability of binary aluminides as well as their mutual solubility. All selected specimens contain Fe, Co and Ni because AlFe, AlCo and AlNi are stable binary aluminides with pronounced mutual miscibility and close lattice constants [29, 30]. Other transition metal elements, *e.g.* Mn, Cu and Cr, were also added into the matrix. In addition, 50% of the Al atoms were replaced with Ti in HEAL #6 and #7 to further perturbate the B2 structure and extend the work beyond pure aluminides.

High-purity powders Al (99.9%), Ti (99.5%), Cr (99.95%), Fe (99.9%), Co (99.8%), Ni (99.9%), Cu (99.9%) and Mn (99.95%) purchased from Alfa Aesar were utilized as starting materials. Appropriate amounts of each powder according to the stoichiometry were used to fabricate specimens. The seven compositions for forming B2-phase HEICs are listed in Table 1 and referred to as specimen #1 to #7 in the text. The raw powders were mechanically alloyed via high energy ball milling (HEBM) using a SPEX 8000D miller (SpexCertPrep, NJ, USA) for 3 hours. To prevent overheating, the HEBM was stopped every 60 minutes to allow cooling for 5 minutes. The powders were then compacted into disks of 20-mm diameter and consolidated by using spark plasma sintering (SPS, Thermal Technologies, CA, USA).

All the consolidated bulk samples were further homogenized at 1100 °C for 10 hours. The phase composition and lattice parameters were determined by XRD using a Rigaku diffractometer with Cu Kα radiation. The compositions of the specimens were characterized by scanning electron microscopy equipped with energy dispersive X-ray spectroscopy (EDXS). Specifically, EDXS mapping was employed to image the size and distribution of the fine precipitates.

To test the reversible temperature-dependent stability of the B2-phase $(Co_{1/4}Fe_{1/4}Ni_{1/4}Cu_{1/4})Al$ was chosen and annealed at 1000 °C, 1100 °C, and 1300 °C, respectively, for 10 hours, in a sequence. Subsequently, the same specimen (after equilibrated at the higher temperature of 1300

°C) was annealed again at the lower temperature 1000 °C isothermally for 10 hours. After each annealing step, the specimen was quenched and XRD measurement was performed to determine the reversible temperature-dependence of the phase stability.

To extend the generality of this study, five additional compositions (with 75% Al and equal molar amounts of four other transition metal elements) were selected to fabricate $D0_{22}$-phase HEICs via the same mechanical alloying and SPS procedure, followed by annealing at 1300 °C.

### 3. Results and Discussion

#### 3.1 The Formation of High-Entropy B2 Phases

The phase formation of the seven specimens equilibrated at 1100 °C was determined by XRD patterns, as shown in Fig. 2(a). For comparison, the diffraction peaks for the pure BCC (Fe as an example) and B2 structured aluminide (FeAl as an example) were also simulated. It is noted that the B2 phase (a BCC-based structure with CsCl ordering) has BCC-like diffraction peaks of (110), (200) and (211), which are indicated by dots in Fig. 2. However, the formation of ordered B2 phase is characterized by the "superlattice" diffraction peaks, (100), (111) and (210), as indicated by stars, which are "forbidden reflections" in the disordered BCC structure due to the symmetry. XRD patterns of specimens #1 to #5 match well with the B2 phase (with virtually no detectable secondary phases), indicating the successful synthesis of the B2-phase high-entropy aluminides (as one type of HEICs).

For the two Ti-contained specimens #6 $(Fe_{1/4}Co_{1/4}Ni_{1/4}Cu_{1/4})(Al_{1/2}Ti_{1/2})$ and #7 $(Fe_{1/4}Co_{1/4}Ni_{1/4}Mn_{1/4})(Al_{1/2}Ti_{1/2})$, XRD patterns showed the formation of very small amounts of secondary precipitate phases.

Derived from the XRD patterns, the lattice parameter obtained for the B2 structure for the specimens #1 to #5 is close to 2.89 Å (within 0.7% variations), as shown in Table 1. The measured lattice parameters for specimens #6 $(Fe_{1/4}Co_{1/4}Ni_{1/4}Cu_{1/4})(Al_{1/2}Ti_{1/2})$ and #7 $(Fe_{1/4}Co_{1/4}Ni_{1/4}Mn_{1/4})(Al_{1/2}Ti_{1/2})$, in which 50% Al atoms were replaced by Ti atoms (with a 15% larger atomic radius), are increased to around 2.95Å.

#### 3.1 Anti-Site Defects in HEALs

Anti-site defects are commonly present in the intermetallic compounds at finite temperatures due to an entropic effect [31]. In a conventional B2 intermetallic compound AB, four possible types of substitutional point defects (anti-site A and B atoms, as well as two types of vacancies in two sublattices) can exist [31], and there are more compositional variables of the anti-site defects in HEICs.

Using the single-phase specimen #3 $(Co_{1/4}Fe_{1/4}Ni_{1/4}Cu_{1/4})Al$ as an example, we simulated a series of XRD patterns to estimate the anti-site occupation assuming, for simplicity, equimolar transition metal anti-site defects in the Al sublattice and ignoring vacancies. The modeled relative intensity of the (100) superlattice peak (normalized to the strongest (110) peak) *vs.* anti-site occupation fraction is plotted in Fig. 2(b). The anti-site defects were then estimated to be ~10% by comparing the measured relative intensities with the simulation results. Other specimens (#1, #2, #3, #4, and #5) are estimated to have similar levels of anti-site defects at 1100 °C.

### 3.3 Compositional Homogeneity and Single-Phase HEICs

To examine the compositional homogeneity, SEM and EDXS were performed. The EDXS elemental mapping showed that the specimens #3 $(Co_{1/4}Fe_{1/4}Ni_{1/4}Cu_{1/4})Al$ and #5 $(Co_{1/5}Fe_{1/5}Ni_{1/5}Mn_{1/5}Cu_{1/5})Al$ are compositionally homogeneous without any detectable secondary phase.

However, Cr-enriched regions in specimens #1 $(Co_{1/4}Fe_{1/4}Ni_{1/4}Cr_{1/4})Al$ and #4 $(Co_{1/5}Fe_{1/5}Ni_{1/5}Mn_{1/5}Cr_{1/5})Al$ were revealed by EDXS mapping, despite they both appear to be single phases in XRD. EDXS mapping also suggests that specimen #2 $(Co_{1/4}Fe_{1/4}Ni_{1/4}Mn_{1/4})Al$ has very small amount of Mn-enriched precipitates. In additional, EDXS showed that specimen #6 $(Co_{1/4}Fe_{1/4}Ni_{1/4}Cu_{1/4})(Al_{1/2}Ti_{1/2})$ had Cu-enriched precipitates, and specimen #7 $(Fe_{1/4}Co_{1/4}Ni_{1/4}Mn_{1/4})(Al_{1/2}Ti_{1/2})$ contained a secondary phase enriched with Ti, Fe, and Mn.

In summary, the combination of XRD and EDXS compositional mapping (Fig. 2) showed that specimens #3 $(Co_{1/4}Fe_{1/4}Ni_{1/4}Cu_{1/4})Al$ and #5 $(Co_{1/5}Fe_{1/5}Ni_{1/5}Mn_{1/5}Cu_{1/5})Al$ indeed exhibit single, high-entropy B2 phases that are compositionally homogenous. The other five specimens consist of primarily single high-entropy B2 phases, while small amounts of secondary phases or inhomogeneous regions exist; specifically, specimens #1, #2, and #4 appear to be single B2-phase HEICs in XRD patterns while EDXS mapping revealed compositional inhomogeneity.

### 3.4 CALPHAD Modeling

We further conducted CALPHAD (calculation of phase diagram) modeling to assess the predicted phase fraction *vs.* temperature curves from the ThermoCalc TCHEA database for these seven specimens (Fig. 4). While we recognize none of the current databases have been validated for HEICs (as this is the first experimental study to synthesize single B2-phase high-entropy aluminides), some useful trends can be still obtained, albeit not completely accurate, from the existing databases such as the TCHEA based on the extrapolation from the binary and ternary systems.

The CALPHAD results showed that the primary phase at high temperatures (around 1100 °C) is ordered B2 phase for all samples. Among them, the B2 phase fractions of the specimens #2, #3 and #5 are high (nearly 100%) in the temperature range of 1000 to 1400 K, which are consistent with experiments. The CALPHAD results also indicated that both specimens #1 and #4 should contain a Cr-rich secondary BCC phase (with ~55%Cr, ~15% Fe and ~25% Al) and a Fe-Mn-Ti-enriched Laves phase should precipitate in specimen #7, which all agree with the EDXS elemental maps (Fig. 3(c)).

However, the CAPHAD modeling failed to predict the precipitation of the secondary phase in specimen #6, as observed in the experiment. Moreover, CALPHAD predict a single B2 phase from 700K to the solidus temperature ~1600K for specimen #2, but we observed a reversible secondary phase formation at 1100 ºC or lower (Fig. 4). It is not surprising that CALPHAD modeling is not completely accurate, since this current study represents the first experimental effort to synthesize single B2-phase high-entropy aluminides; thus, all current databases are likely extrapolation from the partial data of binary and ternary aluminides without direct confirmation and calibration of the high-entropy B2-phase stability.

**3.5 Entropy-Stabilized Phases**

Although the CALPHAD predictions (presumably extrapolated from the data of binary and ternary aluminides) are not completely accurate, they (with extrapolation from the binary and ternary data) are able show an important (and valid) general trend (Fig. 4): i.e., the equimolar, single high-entropy B2 phases are generally stable at high temperatures (just below the solidus temperatures), while some secondary phases precipitate at lower temperatures. This general trend suggests that these B2-phase HEICs are entropy-stabilized.

To more critically assess this hypothesis of entropy stabilization, specimen #2 ($Co_{1/4}Fe_{1/4}Ni_{1/4}Mn_{1/4}$)Al was selected to study the reversible temperature stability of the high-entropy B2 phase. The XRD patterns in Fig. 4 showed that specimen #2 exhibited two phases, the primary B2 phase and a secondary phase indicated by solid triangles, when equilibrated at 1000 ºC. The amount of the secondary phase reduced after further annealing at 1100 ºC, and the secondary phase vanished (dissolved completely into the B2 matrix) at 1300 ºC to form a single high-entropy B2 phase. Interestingly, when the same sample was isothermally annealed again at 1000 ºC after forming single high-entropy B2 phase at 1300 ºC, the secondary phase emerged again. The EDXS maps shown in Fig. 4(b) further confirmed the precipitation at 1000 ºC, and the dissolution of the precipitates at 1300 ºC. Specifically, the precipitates are a Mn-rich phase with composition close to 50% Mn, 25% Al, 15% Fe, 5% Co, and 5% Ni.

Similar reversible precipitation of a CuO-enriched secondary phase at low temperatures in $(Mg_{0.2}Co_{0.2}Ni_{0.2}Cu_{0.2}Zn_{0.2})O$ was reported previously and considered as the main evidence of forming this entropy-stabilized oxide [9]. Thus, Fig. 4 also implies that this single-phase equimolar HEIC observed in this study is likely entropy-stabilized at high temperatures.

It is also interesting to note that while specimen #5 $(Co_{1/5}Fe_{1/5}Ni_{1/5}Mn_{1/5}Cu_{1/5})Al$ exhibits a single high-entropy B2 phase at 1100 °C, six out of the ten equimolar ternary aluminide subsystems, i.e., $(Co_{1/2}Ni_{1/2})Al$, $(Cu_{1/2}Co_{1/2})Al$, $(Mn_{1/2}Cu_{1/2})Al$, $(Fe_{1/2}Cu_{1/2})Al$, $(Ni_{1/2}Mn_{1/2})Al$, and $(Fe_{1/2}Mn_{1/2})Al$, do not form single B2 phases at the same temperature (based on CALPHAD modeling, which should be more accurate for ternary subsystems). This also suggests a high-entropy effect to stabilize equimolar solid solutions in high-component-dimensional systems.

### 3.6 Criterion for Forming Single High-Entropy B2 Phases

Two criteria are commonly used to assess the formation of single-phase HEAs, i.e., atomic size polydispersity $\delta = 100\sqrt{\sum_{i=1}^{n} c_i(1 - r_i/\bar{r})^2}$ and mixing enthalpy $\Delta H_{mix} = \sum_{i=1, i \neq j}^{n} c_i c_j \Omega_{ij}$, where the average atomic radius $\bar{r} = \sum_{i=1}^{n} c_i r_i$, $c_i$ and $r_i$ are the atomic percentage and atomic radius of the $i^{th}$ element, respectively, $n$ is the number of alloying elements, and $\Omega_{ij}$ is the mixing enthalpy between $i$ and $j$ in the liquid phase [32]. It is proposed that the formation of a single-phase HEA solid solution is favored when $\delta \leq 6$ and $-15$ kJ/mol $\leq \Delta H_{mix} \leq 5$ kJ/mol. An additional indicator, VEC = $\sum_{i=1}^{n} c_i \chi_i$, where $\chi_i$ is the valence electron concentration (VEC), was introduced to predict whether an HEA forms BCC (VEC < 6.5) or FCC (VEC > 6.5) structure. [33]. The corresponding values of $\delta$, $\Delta H_{mix}$, and VEC for seven specimens were calculated and listed in Table 1. Here, small VEC values (2.6-3.5) suggest they would favor to form BCC-like structure, but large $\delta$ values (> 7 for all cases) show that they should not form HEAs of the simple BCC structure. The predictions are consistent with the experimental observations that they form BCC-based, ordered B2 phases.

The atomic size polydispersity $\delta$ calculated with 50% Al should be used to judge whether an HEA solid solution on one lattice (simple FCC or BCC) can to form. The ordered B2 HEICs are more like HECs with one element (Al in this case vs. anions in HECs) primarily occupies one sublattice and other four or five elements form a solid solution on the other sublattice. In HECs, atomic size difference of cation atoms and cation-anion bonding lengths are considered as a key parameter for predicting the formation of single HEC phases [10, 11]. Here, we proposed to use atomic size polydispersity of non-Al elements, $\delta^* = 100\sqrt{\sum_{i \neq Al} c_i(1 - r_i/\bar{r}_{no-Al})^2}$, as a refined parameter to the formation of single high-entropy B2 phases. The calculated results are listed in Table 1. It is found that the specimens #3 and #5 (single HEIC phases at 1100 °C) as well as specimen #2 (almost single high-entropy B2 phase at 1100 °C and true single high-entropy B2 phase at a high temperature of 1300 °C ) with $\delta^* < 1$ form single high-entropy B2 phases, thereby suggesting that $\delta^*$ could be a good indicator for forming single-phase HEICs.

A careful examination further shows that the calculated $\Delta H_{mix}$ values are in the range of $-12$ to $-29$ kJ/mol. It is interesting to note that only specimens #3 and #5 have $\Delta H_{mix} > -15$ kJ/mol, which are the two that form single high-entropy B2 phases at 1100 °C. For ordered B2-phase HEICs, perhaps it is more accurate to use the sum of average weighted bonding energies between non-Al and Al atoms to define a modify $\Delta H_{mix}^* = \sum_{i \neq Al} c_i \Omega_{Al-i}$. The calculated results are listed in Table 1. Again, only specimens #3 and #5, the two that form single high-entropy B2 phases at 1100 °C, have the refined $\Delta H_{mix}^* > -15$ kJ/mol and $\delta^* < 1$.

It is interesting to note that specimen #2 has the lowest $\delta^*$ of 0.66, but with a more negative $\Delta H_{mix}^*$ of $-17.75$ kJ/mol. Experimentally, only trace amount of secondary phase was observed in specimen #2 at 1100 °C (Fig. 2); moreover, it formed single high-entropy B2 phase at a higher temperature of 1300 °C, but a secondary phase reversibly precipitated out at a lower temperature of 1000 °C. Presumably, the negative $\Delta H_{mix}^*$ effect was offset by higher entropy stabilization effect at a higher temperature, which further supports that this single high-entropy B2 phase is entropy-stabilized at high temperature.

Overall, we conclude that a small $\delta^*(< 1)$ appears to be a good indicator to promote the formation of a single high-entropy B2 phase (and perhaps HEICs in general), while a less negative $\Delta H_{mix}^*$ may help to stabilize HEICs to a lower temperature. More critical testing and possible refinement of the proposed criterion should be carried out in future studies.

**3.7 High-Entropy D0$_{22}$ Phases**

To extend the generality of this study and our discovery of single-phase HEICs, we further examined the possible formation of D0$_{22}$-phase HEICs (see Fig. 1(b) for the schematic structure) fabricated via the same route. We selected possible compositions based on the following principles (similar to those used for selecting possible B2-phase HEICs): (1) at least three transition metal elements are able to form equilibrium D0$_{22}$-phase binary aluminides and (2) at least two of the D0$_{22}$-phase binary aluminides have 100% mutual solubility according to the ternary phase diagrams. Fig. 5 shows the XRD patterns of the five specimens of selected compositions: i.e., $(Ti_{1/4}Nb_{1/4}V_{1/4}Zr_{1/4})Al_3$, $(Ti_{1/4}Nb_{1/4}Ta_{1/4}Cr_{1/4})Al_3$, $(Ti_{1/4}Nb_{1/4}Ta_{1/4}Mn_{1/4})Al_3$, $(Ti_{1/4}Nb_{1/4}Ta_{1/4}Mo_{1/4})Al_3$, and $(Ti_{1/4}Nb_{1/4}Ta_{1/4}Zr_{1/4})Al_3$. All these five compositions exhibited the primary $D0_{22}$ phases after annealing at 1300 °C for 10 hrs. Very few amounts of secondary phases were detected by XRD measurements. These findings further demonstrated that equimolar HEICs of mostly single high-entropy phases (albeit small amounts of secondary phases) beyond the high-entropy B2 phases can be made.

### 4. Conclusions

In this study, we successfully fabricated, for the first time to our knowledge, several single-phase HEICs with ordered B2 structure, in which four or five transition metal elements, *e.g.* Fe, Co, Ni, Mn, and Cu, of equimolar fractions, occupy one sublattice, with Al on the other sublattice (albeit ~10% anti-site defects). Specifically, $(Co_{1/4}Fe_{1/4}Ni_{1/4}Mn_{1/4})Al$, $(Co_{1/4}Fe_{1/4}Ni_{1/4}Cu_{1/4})Al$, and $(Co_{1/5}Fe_{1/5}Ni_{1/5}Mn_{1/5}Cu_{1/5})Al$ can be made into single high-entropy B2 phases, while four other compositions exhibit predominantly single high-entropy B2 phases with small amounts of secondary phases. These high-entropy B2 phases are likely entropy-stabilized phases based on CALPHAD modeling and a model experiment. A new criterion for forming single high-entropy B2 phases is proposed as a new phase stability rule. Five additional HEICs of primarily $D0_{22}$ phases have been made to broaden the discovery.

The discovery of single-phase HEICs bridges the traditional metallic HEAs and emerging non-metallic HECs; their structures are more like ionic HECs with high-entropy mixing only on one sublattice. However, comparison of experimental and calculated XRD patterns suggest the existence of ~10% anti-site defects in B2-pgase HEICs (differing from most HECs with little anti-site defects). The single-phase HEICs reported in this study represents a new class of high entropy materials, which opens a new platform to explore unique mechanical, thermal, and other functional properties.

**Acknowledgement**

This work is partially supported by a Vannevar Bush Faculty Fellowship sponsored by the Basic Research Office of the Assistant Secretary of Defense for Research and Engineering and funded by the Office of Naval Research through Grant no. N00014-16-1-2569. T.H. also acknowledges the funding support from State Key Laboratory of High Performance and Complex Manufacturing at Central South University (Grant no. ZZYJKT2018-04).

**Disclose of potential conflict of interest:** The authors declare no conflict of interests.

**Figure Legends**

**Fig. 1.** Schematic illustrations of the high-entropy aluminides with the B2 and $D0_{22}$ structures.

**Fig. 2. (a)** XRD patterns for seven HEIC specimens that exhibit primarily or completely single high-entropy B2 phases after annealing at 1100 ºC for 10 hours. Simulated XRD peaks for the BCC (using Fe as an example), perfectly-ordered B2 (using binary FeAl as an example), and B2 phase with 10% anti-site defects are also included. Peaks that can belong to either BCC or B2 structure are indexed by dots; superlattice peaks belong exclusively to the B2 structure are indexed by stars. Minor peaks are evident in several patterns, e.g., for specimen #6 and #7, which indicate the presence of secondary phases. **(b)** Simulated intensity for (001) peak (normalized to the strongest (110) peak) in B2-structured aluminide as a function of the fraction of anti-site Al defects. Calculation was performed by using VESTA software. **(c)** SEM micrographs and corresponding EDXS compositional maps for specimens after annealing at 1100 ºC for 10 hours. HEIC specimen #3 and #5 appear to completely homogeneous, while specimen #2 were almost homogeneous.

**Fig. 3.** Phase evolution (volume fraction of various phases *vs.* equilibrium temperature) predicted by ThermoCalc using TCHEA database. The CALPHAD can be used to forecast some general trends (to the first order of approximation), but the predictions are not all accurate when they are compared with Fig. 2 and Fig. 4; this is presumably due to that the database does not include all interactions in aluminides.

**Fig. 4 (a)** XRD patterns of the same #2 $(Fe_{1/4}Co_{1/4}Ni_{1/4}Mn_{1/4})Al$ specimen annealed at 1000 ºC, 1100 ºC, 1300 ºC, and 1000 ºC sequentially (each for 10 hours). **(b)** The enlarged XRD peaks showing the evolution of the secondary phase. **(c)** EDXS mapping of the same specimen annealed at 1300 ºC and 1100 ºC, respectively. The secondary phase that formed at 1000 ºC and 1100 ºC vanished after annealing at high temperature of 1300 ºC, but re-precipitated after subsequent re-annealing at 1000 ºC. This model experiment implies that the single B2 solid-solution phase is likely entropy-stabilized at high temperatures (and the CALPHAD prediction from the TCHEA database shown in Fig. 3(b) is not all accurate).

**Fig. 5.** XRD patterns of five HEIC specimens with primarily the high-entropy $D0_{22}$ phase after annealing at 1300 ºC for 10 hours. The $D0_{22}$ phase are indexed; the unindexed peaks with low intensity correspond to the secondary phases. The high-entropy $D0_{22}$ phase dominant in all five cases (albeit some minor secondary phases).

**Table 1.** Summary of the seven specimens that exhibit primarily or completely single high-entropy B2 phases after equilibration at 1100 °C for 10 hours.

|    | Composition | Single B2 Phase at 1100 °C? | Lattice Constant (Å) | $\delta$ | $\Delta H_{mix}$ | VEC | $\delta^*$ | $\Delta H_{mix}^*$ |
|----|-------------|------------------------------|----------------------|----------|------------------|-----|------------|---------------------|
| #1 | $(Fe_{1/4}Co_{1/4}Ni_{1/4}Cr_{1/4})Al$ | No | 2.887 | 7.64 | -16.44 | 2.83 | 1.30 | -15.50 |
| #2 | $(Fe_{1/4}Co_{1/4}Ni_{1/4}Mn_{1/4})Al$ | Almost | 2.895 | 7.78 | -18.75 | 2.86 | 0.66 | -17.75 |
| #3 | $(Fe_{1/4}Co_{1/4}Ni_{1/4}Cu_{1/4})Al$ | Yes | 2.918 | 7.72 | -12.00 | 3.34 | 0.97 | -13.25 |
| #4 | $(Fe_{1/5}Co_{1/5}Ni_{1/5}Mn_{1/5}Cr_{1/5})Al$ | No | 2.909 | 7.57 | -17.24 | 2.69 | 1.18 | -16.20 |
| #5 | $(Fe_{1/5}Co_{1/5}Ni_{1/5}Mn_{1/5}Cu_{1/5})Al$ | Yes | 2.916 | 7.64 | -13.96 | 3.16 | 0.92 | -14.40 |
| #6 | $(Fe_{1/4}Co_{1/4}Ni_{1/4}Cu_{1/4})(Al_{1/2}Ti_{1/2})$ | No | 2.948 | 7.12 | -24.00 | 3.12 | / | / |
| #7 | $(Fe_{1/4}Co_{1/4}Ni_{1/4}Mn_{1/4})(Al_{1/2}Ti_{1/2})$ | No | 2.945 | 7.20 | -28.38 | 2.65 | / | / |

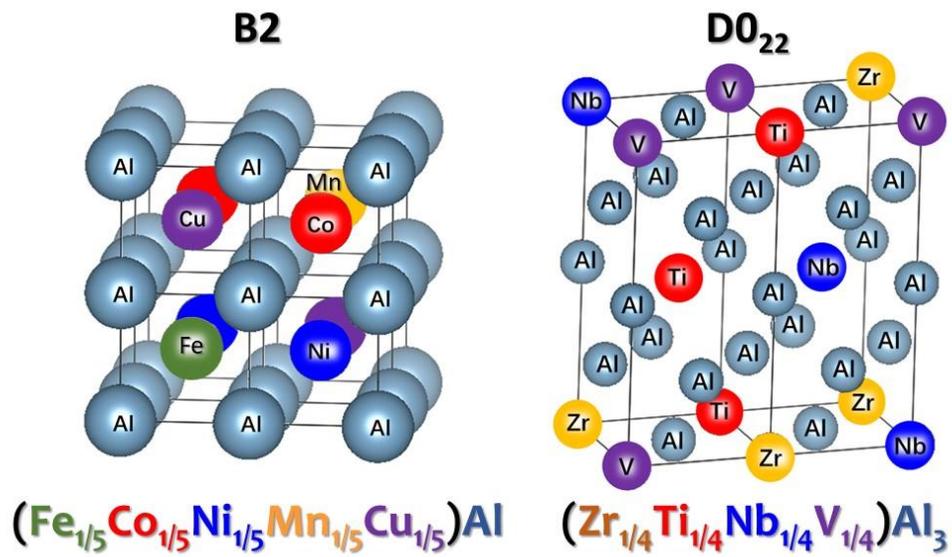

**Fig. 1.** Schematic illustrations of the high-entropy aluminides with the B2 and D0$_{22}$ structures.

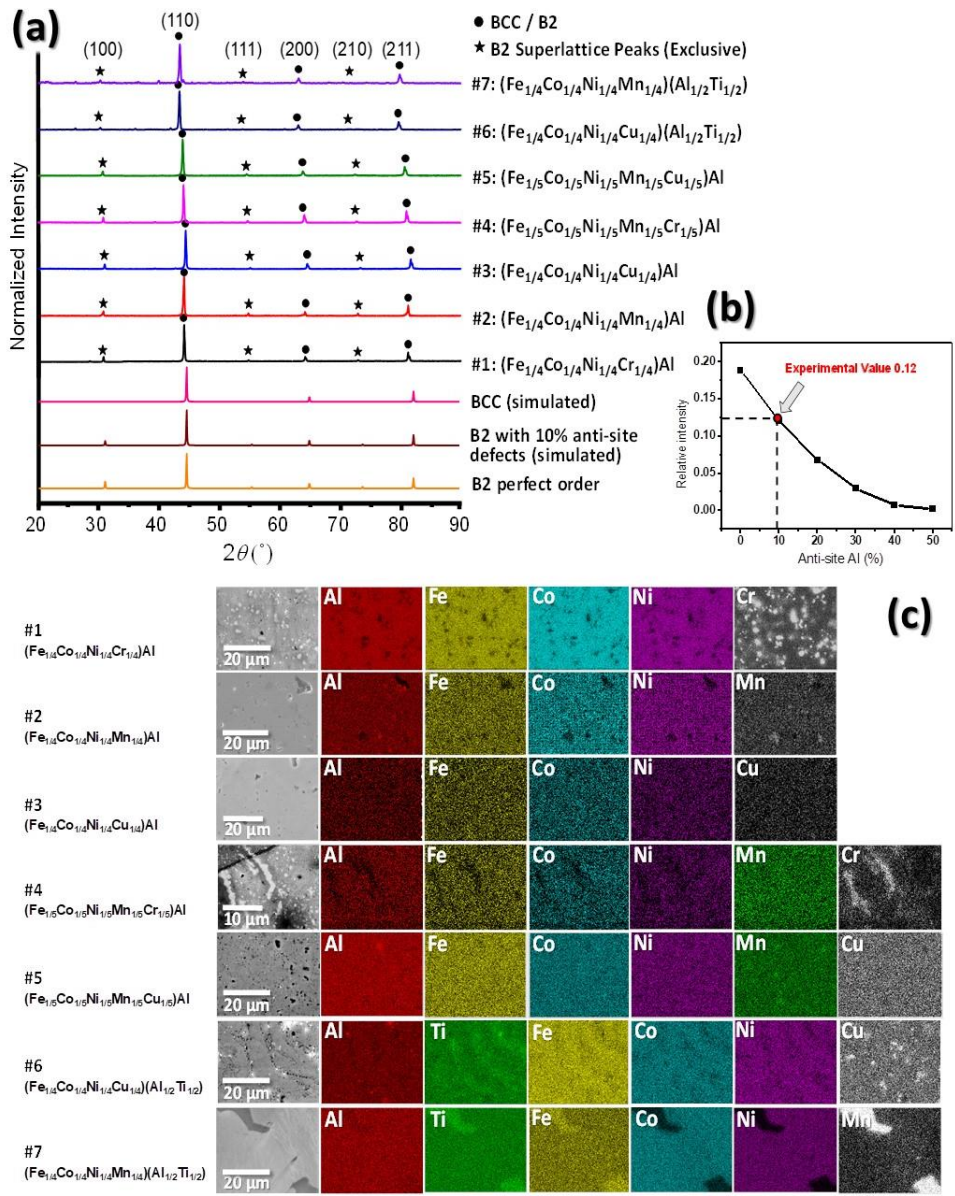

**Fig. 2. (a)** XRD patterns for seven HEIC specimens that exhibit primarily or completely single high-entropy B2 phases after annealing at 1100 °C for 10 hours. Simulated XRD peaks for the BCC (using Fe as an example), perfectly-ordered B2 (using binary FeAl as an example), and B2 phase with 10% anti-site defects are also included. Peaks that can belong to either BCC or B2 structure are indexed by dots; superlattice peaks belong exclusively to the B2 structure are indexed by stars. Minor peaks are evident in several patterns, e.g., for specimen #6 and #7, which indicate the presence of secondary phases. **(b)** Simulated intensity for (001) peak (normalized to the strongest (110) peak) in B2-structured aluminide as a function of the fraction of anti-site Al defects. Calculation was performed by using VESTA software. **(c)** SEM micrographs and corresponding EDXS compositional maps for specimens after annealing at 1100 °C for 10 hours. HEIC specimen #3 and #5 appear to completely homogeneous, while specimen #2 were almost homogeneous.

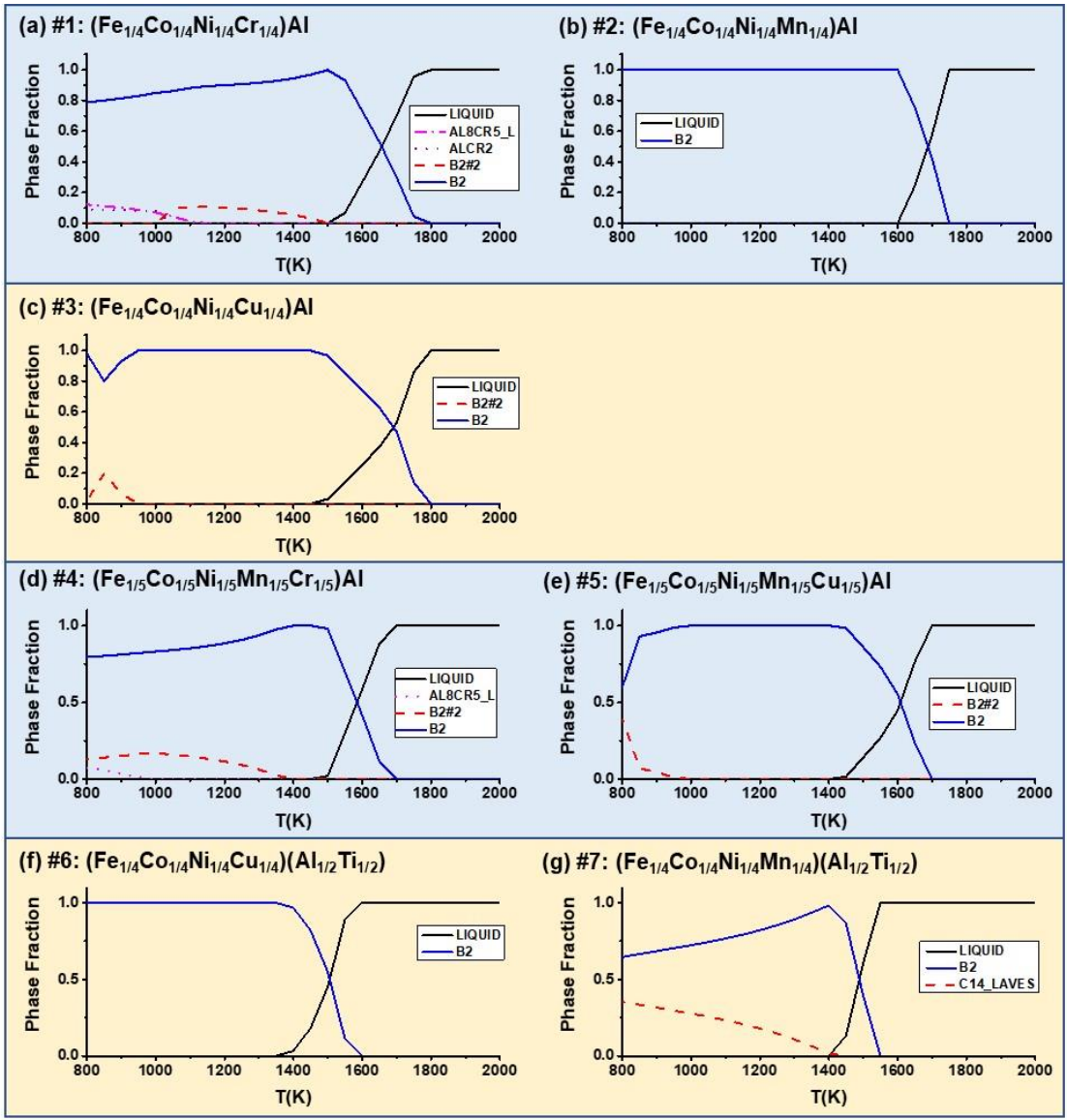

**Fig. 3.** Phase evolution (volume fraction of various phases *vs.* equilibrium temperature) predicted by ThermoCalc using TCHEA database. The CALPHAD can be used to forecast some general trends (to the first order of approximation), but the predictions are not all accurate when they are compared with Fig. 2 and Fig. 4; this is presumably due to that the database does not include all interactions in aluminides.

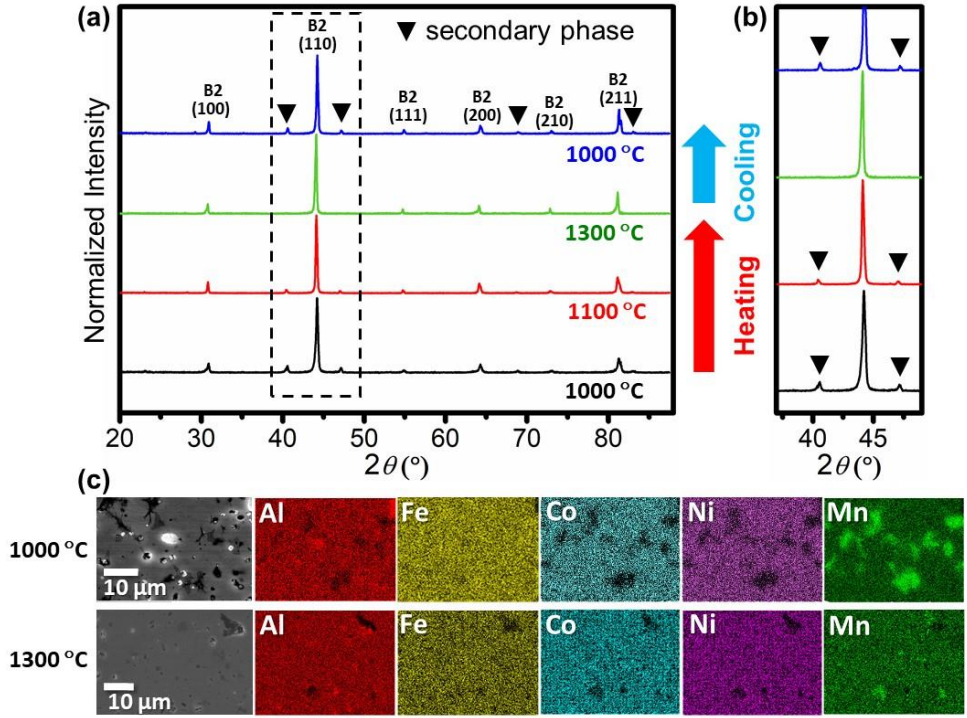

**Fig. 4 (a)** XRD patterns of the same #2 $(Fe_{1/4}Co_{1/4}Ni_{1/4}Mn_{1/4})Al$ specimen annealed at 1000 °C, 1100 °C, 1300 °C, and 1000 °C sequentially (each for 10 hours). **(b)** The enlarged XRD peaks showing the evolution of the secondary phase. **(c)** EDXS mapping of the same specimen annealed at 1300 °C and 1100 °C, respectively. The secondary phase that formed at 1000 °C and 1100 °C vanished after annealing at high temperature of 1300 °C, but re-precipitated after subsequent re-annealing at 1000 °C. This model experiment implies that the single B2 solid-solution phase is likely entropy-stabilized at high temperatures (and the CALPHAD prediction from the TCHEA database shown in Fig. 3(b) is not all accurate).

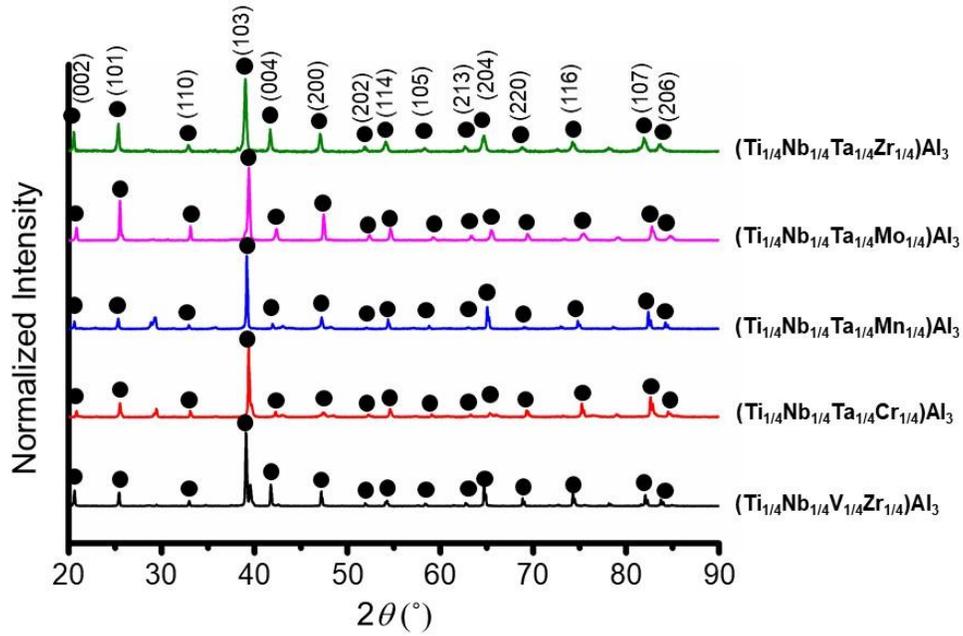

**Fig. 5.** XRD patterns of five HEIC specimens with primarily the high-entropy $D0_{22}$ phase after annealing at 1300 ºC for 10 hours. The $D0_{22}$ phase are indexed; the unindexed peaks with low intensity correspond to the secondary phases. The high-entropy $D0_{22}$ phase dominant in all five cases (albeit some minor secondary phases).